\documentclass[12pt]{article}
\usepackage{graphicx}
\usepackage{amsmath}
\usepackage{hyperref}
\usepackage{geometry}
\usepackage{tabularx}
\usepackage{caption}
\usepackage{array} 
\usepackage{makecell} 
\usepackage{booktabs}
\begin{document}

\title{Evaluating the Predictive Power of Qualifying Performance in Formula One Grand Prix}
\author{
    Joshua Weissbock \\
    Carleton University \\
    \texttt{joshweissbock@cmail.carleton.ca} \\
    \and
    Dr. Shirley Mills \\
    Carleton University \\
    \texttt{shirleymills@cunet.carleton.ca}
}
\maketitle

\begin{abstract}
Formula One race weekends are structured around multiple sessions: practices, qualifying, and the Grand Prix itself, each contributing to final race performance. This study analyzes nearly two decades of races, encompassing 7,800 driver-weekend observations, to quantify the predictive value of each session on race outcomes. Expanding on prior research with a larger dataset and a longitudinal perspective, statistical analyses using contingency coefficients and Ordinal Logistic Regression update prior knowledge that qualifying performance is the strongest determinant of final race position, surpassing race start positions and practice results as indicators. Unlike grid positioning at the start of the race, which is susceptible to penalties and external disruptions, qualifying results provide an unbiased measure of driver and car capability. These findings reinforce the strategic emphasis already placed on qualifying, offering an empirical validation of its dominant role in determining race outcomes across varied seasons and eras.
\end{abstract}


\section{Introduction}
\subsection{Background}
Formula One (F1) is the highest class of international auto-racing for open-wheel, single-seat formula racing cars sanctioned by the Fédération Internationale de l'Automobile (FIA)\cite{FIA2024}. As the pinnacle of motor sport, F1 incorporates the latest in all areas of science and engineering. Governed by FIA, F1 runs 20-25 Grand Prix races over a calendar year leading up to both the Driver and the Constructors World Championships. Each race weekend generally follows a structured format of: 2 practices sessions on a Friday, to allow teams to collect data and home in on their technical setups; a third practice session and a qualification session on Saturday, the latter used to determine fastest to slowest cars to set the starting grid in a single lap time trial session; and a Grand Prix on a Sunday, a race of just over 300 km.

F1’s reliance on cutting-edge technology makes data essential for competitiveness. Teams analyze telemetry, tyre degradation, weather, altitude, and competitor performance to optimize decisions. The immense volume and speed of data processing required exemplify advanced data science integration. From race-day tactics to season-long strategy, data-driven insights are vital for success \cite{Budden2020}. This highlights the importance of capturing and maximizing every race weekend variable, from practice sessions to starting grid positions and final race results, to find competitive advantages where possible.

\subsection{Problem Statement}
The relationship between the starting-grid and finishing positions in an F1 race has long been the study of both teams privately, as well as publicly and academically. Previous studies have analyzed this problem and found a relationship in these studies but were limited in scope and techniques. This study aims to address the limitations of previous work, incorporating a larger data set over a longer period of time, as well as reviewing the evolution of a full weekend and the changes year-over-year. By addressing these shortcomings this paper aims to enable actionable outcomes to help teams ahead of a qualification session (which determines initial grid placement before adjustments due to penalties), and improve the predictive understanding of race performance in an F1 weekend.

\subsection{Literature Review}
\label{sec:litreview}
Historical research within F1 focused on the technical, physiological, and structural factors influencing motorsport performance. \cite{Hughes1968} reviewed motor racing importance within the context of evaluating driver skill and car performance and highlighted the importance of measurement systems. \cite{Schwaberger1987} analyzed the physiological demands on an F1 driver required to manage the intense forces experience during a race, emphasizing the high fitness levels required. \cite{Bisagni2008} looked at the structural optimization of composite in F1 cars demonstrating the critical role of engineering within competitive performance.

Pivoting to analyzing the relationship between the starting grid and race outcomes saw various studies emerge in the past 15 years. \cite{Muhlbauer2010} provided an early investigation to this topic, looking at 70 races from 2006 to 2009. This study is the inspiration for this paper, while it found a strong correlation between the starting grid and finishing positions it has many limitations. It suffers from a small sample size, both in drivers (the paper only looked at the top 8 per race, while there are normally 20+ per race) and season (just 4) - all providing a limited scope of analysis. Additionally, the authors fail to address that the starting grids are already established by F1's qualification process, a time-trial allowing cars to determine which is the fastest in a single lap ahead of the race; intuitively the fastest car on a Saturday should likely be the fastest car on a Sunday which potentially overstates the importance of grid positions.

In addition to this keystone study, \cite{McCarthy2013} reviewed strategic behaviours associated with the starting grid, particularly focusing on their influence in first-turn incidents. \cite{VanKesteren2022} advanced the field by using Bayesian Analysis to try and separate driver-skill and car-performance over multiple races in order to provide a more robust approach to analyzing race results. \cite{Padilla2023} investigated rookie F1 drivers and how much experience plays a role in determining team success. These papers all try and address starting position, driver skill and ultimately performance and outcomes within an F1 race. 

\subsection{Objective and Contributions}
Addressing the gaps identified within \cite{Muhlbauer2010}, this study aims to address these limitations to further evolve our understanding of car performance over a full weekend to maximize performance in the race itself. While these previous studies were valuable as a starting place they suffer from multiple issues this paper aims to address. This work was restricted by both a small sample size of drivers (only top 8 were reviewed while normally 20+ race in a single race) and by a small slice of seasons (2006-2009). Additionally the starting grid is already self sorted placing the fastest cars at the front and this study fails to explore the broader context of a full weekend. Our work leverages an extensive dataset including almost two decades of races to provide a more comprehensive view of F1's evolution over time.

By reviewing all sessions over a weekend, including all practice sessions, qualification and then the race day, we can provide a more holistic analysis to which performance is predictive. This study aims to account for changes in F1 regulation and technology changes over time. 

The results and contributions from this paper are two-fold. First, it provides a more robust framework to how starting grids can be established, and what performance over a weekend is likely to impact it. It also looks at these changes over time. Second, it provides decision makers actionable insights to teams to enable them to refine their strategies and decision making ahead of a qualification session (where results are locked in for the race). This furthers the academic understanding of motorsports strategy, as well as how it fits within the larger practical understanding of competitive strategy.

\section{Materials and Methods}
\subsection{Data Prep and Cleaning}

Data for this study was collected by scraping the result from the official F1 website (\url{http://www.Formula1.com}). This dataset includes all races held from 2006 to 2023, as this scraping process was completed before the end of the 2024 season. In total over 400 races were included and analyzed. Some driver/race information were initially excluded due to their incomplete or disrupted data, primarily when a single chassis would see different drivers for different parts of the weekend (i.e. a single day illness, or allowing rookies to drive for a session). Other Grand Prix were excluded if they were missing sessions, for example the 2023 Emilia Romagna Grand Prix was canceled due to flooding, and the 2017 Chinese Grand Prix experienced severe disruptions to practice sessions caused by adverse weather conditions - these races were removed as they did not contain all 3 days of sessions.

Data cleaning steps were implemented to ensure reliability and consistency of this data. Grands Prix with sprint races, a newer F1 implementation of a mini race (usually half distance) held on the Saturday of a weekend and introduced in 2021, were excluded due to their different weekend structure and weekend approach due to less time to prepare for qualification. Races where the drivers could not participate in all three sessions were removed as well. This could either be because of weather impacting a single session (e.g. United States in 2015 due to rain, Germany in 2020 due to fog or Russia in 2021 due to rain) or because there was a driver change mid-weekend, either due to a rookie having an opportunity to drive an F1 car or due to sickness. This ensured that this data was able to be compared across various weekends keeping the driver factor steady.

The final cleaned dataset was pivoted so all session results were presented in a single row with all key session details. In total there were over 7,800 driver/race entries representing a complete weekend.  Key variables for each row in the dataset, for each session include:

\begin{itemize}
    \item \textbf{Season}: The year of the race.
    \item \textbf{GP}: (Grand Prix) The name of the Grand Prix.
    \item \textbf{gp\_code}: The unique code F1 gives to that F1 weekend.
    \item \textbf{No}: The driver's number
    \item \textbf{Driver}: The name of the driver participating in the race weekend.
    \item \textbf{Car}: The team or manufacturer the driver represented.
    \item \textbf{Gap}: (For each session) Gaps during each practice session compared to the session leader.
    \item \textbf{Laps}: (For each session) The number of laps completed by the driver during the session.
    \item \textbf{Pos}: (For each session) The final ranking position determined during that session.
    \item \textbf{Time}: (For each session) The best lap time the car achieved in that session
    \item \textbf{Qualifying Time}: The best time the car achieved during each stage of the qualification session
    \item \textbf{Race Start Position}: Adjusted grid position accounting for penalties or disqualifications.
    \item \textbf{Race Finish Position}: Final result of the driver in the race.
    \item \textbf{Status}: Whether the driver finished the race (and how far behind the leader) or did not finish (DNF) and the reason if applicable.    
    \item \textbf{Points}: Points awarded to the driver for their finishing position.
\end{itemize}

The full dataset used for this analysis is available upon request. A snapshot of the dataset is provided in Table~\ref{tab:data_snapshot}, showing a slice of some key columns that can be found within the full dataset.

\begin{table}[h!]
\centering
\small
\begin{tabularx}{\textwidth}{|c|l|X|c|c|c|c|}
\hline
\textbf{Season} & \textbf{GP} & \textbf{Driver}  & \textbf{Car}    & \makecell{Pos \\ (Qualy)} & \makecell{Race \\ Start \\ Position} & \makecell{Race \\ Finish \\ Position} \\ \hline
2006            & Australia   & M. Schumacher    & Ferrari         & 11                   & 10                  & NC              \\ \hline
2006            & Australia   & F. Massa         & Ferrari         & 16                   & 15                  & NC              \\ \hline
2006            & Australia   & R. Barrichello   & Honda           & 17                   & 16                  & 7               \\ \hline
2006            & Australia   & J. Button        & Honda           & 1                    & 1                   & 10              \\ \hline
2006            & Australia   & A. Davidson      & Honda           & NaN                  & NaN                 & NaN             \\ \hline
\end{tabularx}
\caption{Sample of the Dataset Showing Key Columns and Entries.}
\label{tab:data_snapshot}
\end{table}

This data cleaning process ensures that the analysis reflects a single driver driving a single car across an entire weekend and how their results evolve over the weekend in order to review the predictability of final race positions. 

\subsection{Experiment, Analysis, and Evaluation Methods}

As described in section~\ref{sec:litreview}, this study validates the work done by Mühlbauer \cite{Muhlbauer2010} and adds to their methodology and results. The relationship between the starting grid position and finishing position was analyzed using the contingency coefficient (C), same as in the original work. This methods continues to be used to enable continuity. C is used as it was specifically designed to measure the strength of association between two nominal variables (e.g. comparing starting position and finishing position) and is well suited for comparing categorical variables. By using C, the original authors were able to account for the association between these two variables with the limited number of categories.  Replicating what they did, we first limited the analysis to the 2006 to 2009 seasons (inclusive), focusing on the top 8 positions in each race. This allowed us to validate their findings and create a baseline from which we can further explore the other sessions in a Grand Prix weekend as well as fully understand the effectiveness of their approach.

The next step we took was to expand upon their work to fullly understand the impact of the race weekend evolution. Ensuring consistency across race we used the full dataset to include the top 20 positions of each race, as most races consist of 20 drivers, ensuring consistency across races. We also examined the years from 2010 to 2023, electing to start at 2010 as this season saw major technical regulation changes as well as the introduction of the current point system. We also included an additional metric in our analysis with the calculation of Spearman's Rho to measure rank-order correlations. While C captures the strength of nominal associations, Spearman’s Rho provides an alternative by assessing monotonic relationships between ranked variables. This additional measurement gives us another perspective to understand the trends within the sport.

Iterating off of the expanded dataset we review the relationship between not just the starting grid and finishing results, but also the results of each practice session (P1, P2 and P3) as well as qualification results to understand the predictive reliability of these sessions. Further analyzing these year over year we can see if the strength of relationship has changed over time highlighting any potential changes in reliability of the grid, or change within the sport.

The final step in our analysis employed an Ordinal Logistics Regression (OLR) model to quantify the relationship between practice \& qualifying sessions with the final race performance. This method is well suited for our target variable (race finishing position groups) which are ordinal categories with a natural ranking but without equal spacing. Unlike linear regression, which assumes continuous and interval-scaled outcomes, OLR handles ordered categorical responses by model the cumulative log odds of being in a given category or lower. This structure, known as the proportional odds model, assumes that the relationship between these predictor variables to the odds of being in a high-ranked category, remains constant across outcomes. This makes OLR ideal for motor sports, like F1, where positions matter, but gaps between these positions may vary for many reasons (e.g. incident-driven race dynamics, strategy, etc). We validate the proportional odds assumption and proceed to evaluate session contributions using this cumulative logit framework. This approach allows for interpretable modeling of how each practice and qualifying session contributes to a driver's likelihood of finishing in a better position \cite{ucla_olr, singh_olr}.

To ensure robustness of the OLR modeling on this data we conducted: Multicollinearity check (e.g. Variance Inflation Factors) to ensure variables were not overly correlated, standardized regression coefficients ($\beta$ values) to assess the relative importance to each race session, model fit evaluation (McFadden's $R^2$) to understand the explanation of variance in race finishing positions and significance testing (p-values) to see which factors had a meaningful impact on race outcomes.

By using OLR we aim to provide a much deeper understanding of how the different phases of an F1 weekend contribute to final race performance. This allows us to assess the strengths of these relationships and enable further strategic decision making that can be applied within the sport.

All analysis was conducted in Python using Pandas and Matplotlib in Python Notebooks. The combination of replication, expanded metrics, and regression modeling ensures a comprehensive evaluation of the relationship between race weekend variables and final results.

\section{Results}

\subsection{Generalized Results}

Starting with the replication of the 2006-2009 seasons, as seen in Mühlbauer’s methodology for the top 8 positions, we were able to obtain similar results of a C of 0.628 as well as a Spearman's Rho of 0.589  (both $p < 0.001$) confirming their work and a strong relationship between starting and finishing position. Expanding this, for the same years, to the top 20 positions saw these correlations increase to 0.768 and 0.741 respective ($p < 0.001$) which highlights the importance of including all drivers on the grid. The relative frequency (Rel Freq) of maintaining starting positions to the end of the race in the top 8 was 0.558 which decreases to 0.543 when the full grid is included.

Post 2010, following regulatory changes in F1 (such as the removal of in-race refueling and expansion of points), the correlation of starting position to race finishing declines. For the top 20 positions in 2010-2023, C was 0.734 and Spearman's Rho was 0.748  ($p < 0.001$). The Rel Freq in this period was 0.516, which suggests a small reduction in consistency from the 2006-2009 eras. Despite these changes we can see Qualification is a better predictor of race finish ($C=0.741$, Spearman's Rho of $0.763$, $p < 0.001$). The Rel Freq of all positions during this era was 0.518 continuing to reflect the critical importance of qualifying success. 

We hypothesize that qualification positions are more predictive of race results, compared to initial starting grid, due to the fact that the qualification process already sorts out the fastest to slowest cars. The starting grid adjusts these results due to the application of penalties (e.g. replacing power units, exceeding gearbox allocations, or infractions during practice) which gives a distorted view of the fastest to slowest cars. This can artificially inflate slower cars, dampen faster cars and distort the observed predictability of starting position. By contrast, these qualification results are more likely to be unadjusted results of outright single lap pace and performance for all cars and drivers in the same conditions, making it a more reliable performance of race outcomes. Teams are encouraged to optimize qualification setup for raw speed, unencumbered by strategic influences and have their drivers start in an optimal position given the difficulty of passing in modern F1. 

Practice sessions also showed various degrees of predictivity with Practice 3 (the third session right before Qualification) having the strongest correlation to race results ($C = 0.674$, Spearman’s Rho $= 0.676$, $p < 0.001$). The Rel Freq for P3 translates into race finishes at 0.350, while P2 is 0.324 and P1 is 0.300 reflecting how teams get more refined closer to the race in the weekend. 

The results are summarized in Table~\ref{tab:correlations}, which outlines the contingency coefficients ($C$), Spearman’s Rho values, and Relative Frequency for various rank comparisons across the analyzed seasons and permutations.

\begin{table}[h!]
\centering
\captionsetup{skip=10pt}
\small
\begin{tabularx}{\textwidth}{|c|c|c|X|X|c|c|c|}
\hline
\textbf{Start} & \textbf{End} & \textbf{Top n} & \textbf{Rank 1}         & \textbf{Rank 2}          & \textbf{C}   & \textbf{Spearman Rho} & \textbf{Rel Freq} \\ \hline
2006           & 2009         & 8              & Race Starting Pos       & Race Finish Pos          & 0.628        & 0.589                 & 0.558            \\ \hline
2006           & 2009         & 20             & Race Starting Pos       & Race Finish Pos          & 0.768        & 0.741                 & 0.543            \\ \hline
2010           & 2023         & 20             & Race Starting Pos       & Race Finish Pos          & 0.734        & 0.748                 & 0.516            \\ \hline
2010           & 2023         & 20             & Qualification Pos       & Race Finish Pos          & 0.741        & 0.763                 & 0.518            \\ \hline
2010           & 2023         & 20             & P3 Position             & Race Finish Pos          & 0.674        & 0.676                 & 0.350            \\ \hline
2010           & 2023         & 20             & P2 Position             & Race Finish Pos          & 0.664        & 0.674                 & 0.324            \\ \hline
2010           & 2023         & 20             & P1 Position             & Race Finish Pos          & 0.627        & 0.629                 & 0.300            \\ \hline
2010           & 2023         & 20             & P3 Position             & Qualification Pos        & 0.742        & 0.757                 & 0.390            \\ \hline
\end{tabularx}
\caption{Correlations between race weekend variables and final race outcomes.}
\label{tab:correlations}
\end{table}

\subsection{Coefficiency Table}

\begin{figure}[htb!]
    \centering
    \includegraphics[width=\textwidth]{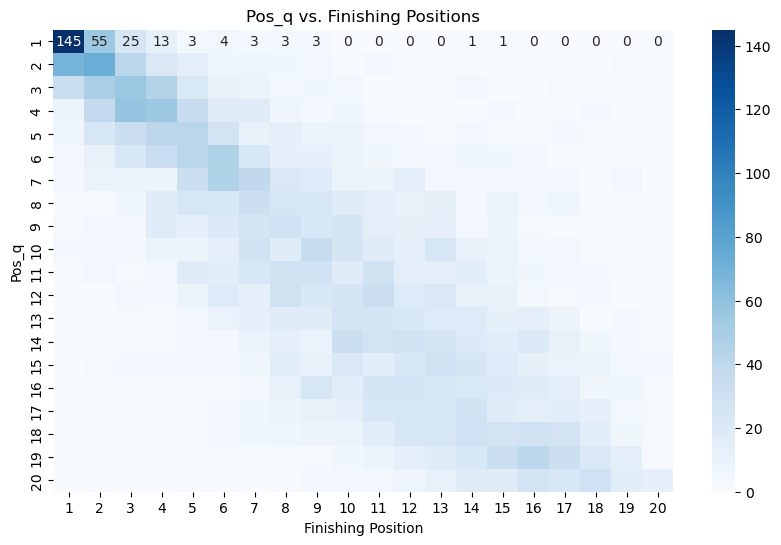}
    \caption{Heatmap showing the relationship between qualifying positions (Pos\_q) and race finishing positions (2010--2023).}
    \label{fig:qual_vs_finish}
\end{figure}

Figure~\ref{fig:qual_vs_finish} presents a heatmap depicting the relationship between qualifying positions (Pos\_q) and race positions for the 2010-2023 season. There is a strong diagonal trend indicating that drivers starting closer to the front tend to finish higher up, and in the points, in the race final. This trend is supported by the data we see above in Table~\ref{tab:correlations}.

Interestingly this relationship tends to breakdown towards the lower-ranked qualifying cars suggesting a less consistent relationship at the back of the grid. While pole position (first on the starting grid, usually first in qualifying) overwhelmingly translates to top race finishes, we see this spread out in the mid-field and back-markers on the grid. This is likely due to the greater susceptibility of these drivers due to first-corner incidents, strategic constraints, overtaking challenges related to DRS (Drag-Reduction System) and moving up in positions when cars ahead of them do not finish due to incidents and reliability.

Overall this suggests that qualifying positions, being unaffected by race-day penalties, is a better indicator of race performance than starting grid positions.

\subsection{Analysis by Eras}

When we examine F1 from the lens of their major technical eras (as seen Table~\ref{tab:technical_eras}, to see how changes in rules might impact this relationship, the relationship between qualifying and finishing positions shows a continued trend. From 2006-2008 during the V8 era, the correlation was moderate ($C = 0.628$, $\rho = 0.589$, $p < 0.001$) with a Rel Freq of 0.557. This improves in the 2009-2013 era where cars used V8s with KERS (Kenetic Energy Recovery System) which saw qualifying become more predictable ($C = 0.744$, $\rho = 0.745$) but Rel Freq dropped to 0.478 likely to reduced reliability of new technologies.

The 2014-2022 turbo-hybrid era demonstrated the strongest correlation between qualifying and race finish ($C = 0.756$, $\rho = 0.777$, $p < 0.001$) with a relative frequency of 0.519. The introduction of hybrid power units further reinforced the importance of qualifying and engineering reliability, as power unit efficiency and durability became key factors in race performance. These trends underscores how F1's evolving technology has shaped race predictability and performance outcomes with qualifying continuing to be the most important moment of performance all weekend.


\begin{table}[htb!]
\centering
\captionsetup{skip=10pt}
\small
\renewcommand{\arraystretch}{1.2} 
\setlength{\tabcolsep}{3pt} 

\resizebox{\textwidth}{!}{ 
\begin{tabular}{|c|c|p{3.2cm}|p{2.8cm}|p{2.8cm}|c|c|c|}
\hline
\textbf{Start} & \textbf{End} & \textbf{Era} & \textbf{Rank 1} & \textbf{Rank 2} & \textbf{C} & \textbf{Spearman’s $\rho$} & \textbf{Rel. Freq} \\ \hline
2006 & 2008 & V10/V8 Engines & Qual. Pos. & Finish Pos. & 0.628 & 0.589 & 0.558 \\ \hline
2009 & 2013 & V8 + KERS & Qual. Pos. & Finish Pos. & 0.744 & 0.745 & 0.478 \\ \hline
2014 & 2022 & Turbo-Hybrid & Qual. Pos. & Finish Pos. & 0.756 & 0.777 & 0.519 \\ \hline
\end{tabular}
} 

\caption{Correlations between qualification and race finish across F1 technical eras.}
\label{tab:technical_eras}
\end{table}

\subsection{Season by Season Analysis}

\begin{figure}[h!]
    \centering
    \includegraphics[width=\textwidth]{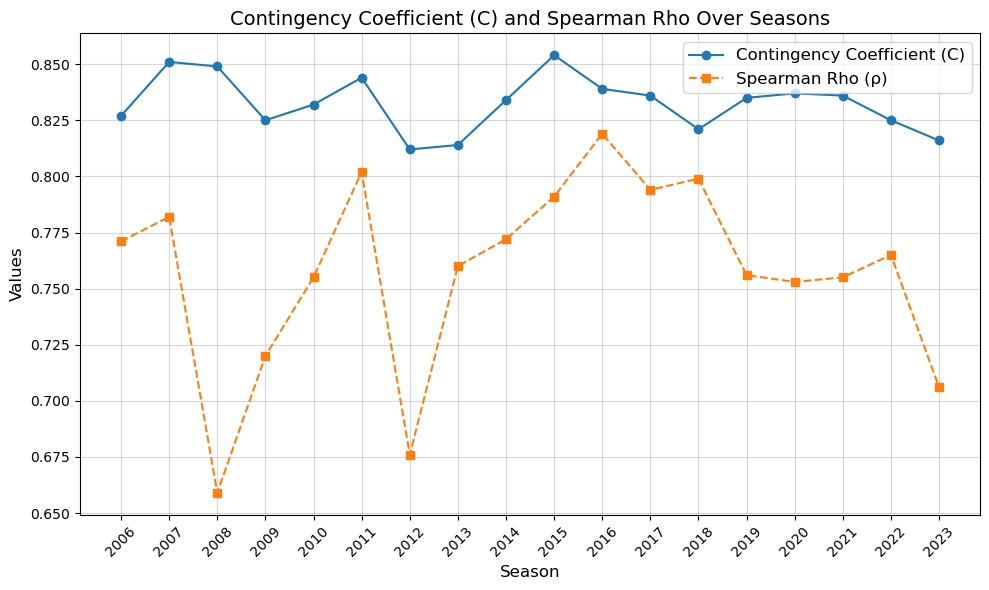}
    \caption{Season-by-season analysis of the contingency coefficient ($C$) and Spearman’s Rho ($\rho$) for qualifying positions vs. race finishing positions (2006--2023).}
    \label{fig:candrhovsseasons}
\end{figure}

Figure~\ref{fig:candrhovsseasons} shows C and Spearman's Rho between qualifying results and final race positions, season by season, for all seasons in the dataset. (Note: while the Y-axis starts at 0.65, these correlation values run from -1 to 1). This figure shows consistent strong correlations across all seasons with $C$ values exceeding 0.81 and $\rho$ over 0.70. There are some notable peaks with $C$ reaching as high as 0.851 (2007) and 0.854 (2015) suggesting especially strong seasons with strong predictability in these values. It suggests that race results were highly influenced by qualifying performance likely due to dominant teams (Ferrari and Mercedes respectively) and due to less external variability such as weather or penalties.

The high values of $C$ and Spearman's Rho in a single season is notably much higher than the correlations previously observed over multiple seasons. This can arise as single season results are less affected by variability introduced over years such as changes in technical regulations, team dynamics or driver performance. When we analyze a car in a single year we can capture the stability of performance less diluted by external factors. This continues to demonstrate that qualifying is a strong predictor of race outcomes, especially in seasons where regulations and performance converge to create a consistent competitive window.

\subsection{Ordinal Logistic Regression}

\subsubsection{Results and Analysis}

An ordinal logistics regression (OLR) model was used to quantify the impact of the practice and qualification sessions placements on final race position. The OLR model included the independent variables of the three practice session rankings (Pos\_p1, Pos\_p2, Pos\_p3) as well as qualification placement (Pos\_q). Table~\ref{tab:olr_results} presents the outputs from the OLR, highlighting the estimated coefficients ($\beta$), standardized regression coefficient ($\beta$) and the statistical significance ($p$) of each predictor.

\begin{table}[h]
    \centering
    \begin{tabular}{lccc}
        \toprule
        Predictor & Coefficient ($\beta$) & Standardized $\beta$ & p-value \\
        \midrule
        Pos\_q (Qualification Position) & 0.2545 & 1.558 & $<0.001$ \\
        Pos\_p3 (Practice 3 Position) & 0.0610 & 0.379 & $<0.001$ \\
        Pos\_p2 (Practice 2 Position) & 0.0576 & 0.368 & $<0.001$ \\
        Pos\_p1 (Practice 1 Position) & 0.0463 & 0.291 & $<0.001$ \\
        \bottomrule
    \end{tabular}
    \caption{Ordinal Logistic Regression Results}
    \label{tab:olr_results}
\end{table}

The results of the OLR suggest that qualification position is the most significant predictor of final race placement with a coefficient of 0.2545 which implies that for every position improved in Qualification improves the log-odds of achieving a better finishing position in the race by 28.9\% ($e^{0.2545} = 1.289)$. Among the practice sessions, Practice 3 has the highest influence on final race standings, then Practice 2 then Practice 1. This suggests that Practice 3 is the most relevant session for race outcomes and reinforces the importance of teams using this time to optimize car setup ahead of qualification. In contrast, Practice 1 has the smallest effect and confirms that early-weekend sessions are better exploratory sessions for data collection or potentially driver test.

While raw coefficients provide some insight into the direct relationship between these variables and the final race result, they cannot be directly compared due to scale in predictors. To address this we calculated  the standardized regression coefficients (also reported in Table~\ref{tab:olr_results}). These standardized coefficients present similar results to the importance to the different sessions and in addition they confirm that Qualification is about four-times larger of an impact than any of the practice sessions. Teams should be using this knowledge to focus on optimizing one-lap performance for better track position. Among the practice sessions, Practice 3 holds as the most important of the sessions and confirms the importance of establishing optimal setup within this window ahead of the qualification session. While teams already use Practice 3 to fine-tune setups, this work provides empirical evidence that performance in P3 significantly predicts race outcomes, reinforcing the importance of this session.

\subsubsection{Threshold Estimates and Performance Barriers}

\begin{figure}[htbp!]
    \centering
    \includegraphics[width=0.7\textwidth]{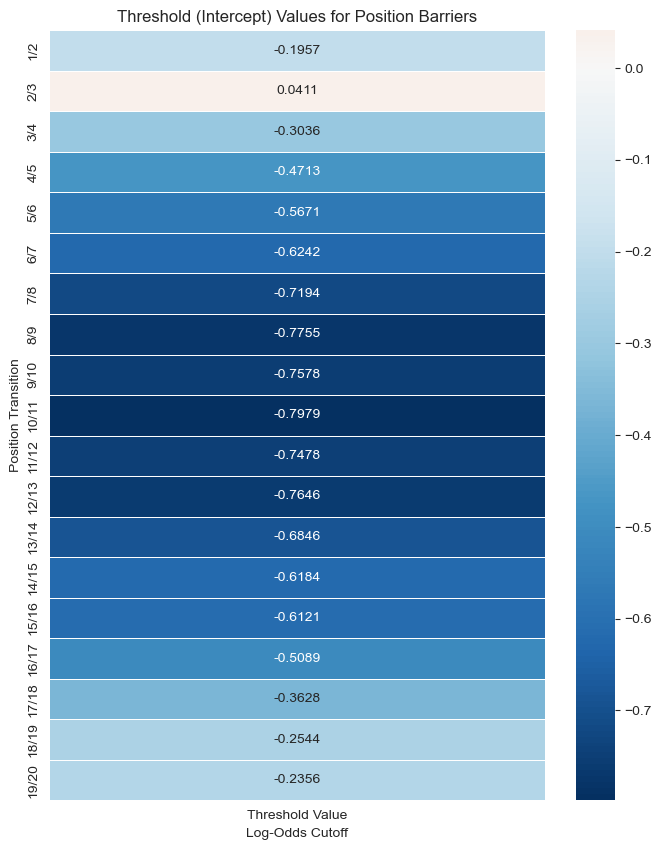}
    \caption{Threshold Estimates (Intercepts) from Ordinal Logistic Regression.}
    \label{fig:olr_thresholds}
\end{figure}

We present the threshold values from the OLR model at Figure~\ref{fig:olr_thresholds}. These thresholds define the log-odds cutoffs between ranking categories which gives us insight into the difficulty of progressing through the grid. Or another way to say this is, these cutoffs represent the log-odds of a driver finishing at or above a given position, the more negative the greater difficulty in surpassing that position while smaller values suggests smoother transitions.

The $1.0/2.0$ threshold ($-0.1957$) is one of the least negative values of all thresholds which suggests that pole position is more vulnerable to being lost than most other transitions (you can't go higher than first). This reflects real world patterns where P1 can be seen to lose the lead in opening laps due to first-corner incidents, bad starts, slip-streaming and strategies. Pole sitters should focus not just on qualification but on their race start procedure.

The $2.0/3.0$ threshold is the most unique threshold as it is the only positive value ($0.0441$) which indicates finishing P2 vs P3 is more evenly distributed compared to other transitions, or to say, once drivers are able to get into the podium finishes, it's much more fluid to swap and finish between P2 and P3. This aligns with real world trends where close margins, race strategy, pit stop execution and team orders can all have an impact.

Conversely, with the $3.0/4.0$ threshold, the negative threshold value ($-0.3036$) indicates a key difficulty in being able to stand on the podium. While P4 is often within reach of a podium finish, P3 needs to maintain a significant performance gap over P4 further emphasizing all the external factors that can affect final race standings.

The most negative threshold value comes from the $10.0/11.0$ threshold ($-0.7979$) which indicates that breaking into the Top 10 is the hardest transition. This aligns with and reflects Formula 1's competitive structure where only the top 10 positions in a race earning points and for backmarker teams, single digit points can make the difference in their final season standings, and thus the earning they can take home at the end of the season. Drivers running in P11 need to consider greater strategic risks to attempt to secure a points finish.

The remainder of the threshold values in the midfield positions tend to indicate a smoother transition indicating that these parts of the field likely see the most frequent position changes. This can come from cars ahead not being able to finish and having an opportunity to move up the ranking, as well as implies these teams have more opportunities to try and move closer to final positions with a point finish.

\subsubsection{Model Evaluation and Fit}

To assess our model's fit, we first computed McFadden's pseudo-$R^2$, which compares the log-likelihood of the fitted model against a null (intercept-only) model. We obtained a pseudo-$R^2 = 0.0166$ indicating that while our model captures directional relationships between session results and final positions, the majority of variance in race outcomes remains unexplained. This reflects the inherently unpredictable nature of Formula One, where race-day incidents, weather, and strategy often override qualifying indicators. The model's Akaike Information Criterion (AIC) of 25,657.28 provides a useful benchmark for comparing competing models. However, this pseudo-$R^2$, as used in ordinal logistic regression, is widely criticized for offering limited insight into the overall suitability of the model \cite{FagerlandHosmer2016, FagerlandHosmer2017}. 

To address this concern we also implemented a Pearson chi-square goodness-of-fit test that compares observed and expected frequencies across the predicted outcome categories. This test yielded a statistic of $\chi^2 = 6,828.25$ with 414 degrees of freedom ($p < 0.001$) which indicates a statistically significant deviation between the predicted and observed outcomes. While this results suggests lack of fit, such chi-square statistics are highly sensitive to large sample sizes and sparse contingency tables which could inflate Type I errors \cite{FagerlandHosmer2016, Lipsitz1996}.

Other options were considered for goodness-of-fit tests, such as the $C_g$ (Hosmer-Lemeshow type), Lipsitz, or Pulkstenis–Robinson (PR) tests, given their difficult in implementation with Python we opted for the Pearson chi-square goodness-of-fit, these others are recommended in literature for their sensitivity to different forms of model misspecification. As an example the $C_g$ and Lipsitz tests are more effective  at identifying issues related to continuous predictors while the PR tests are better suited to models with categorical predictors  \cite{FagerlandHosmer2017, Lipsitz1996, gofcat2023}. Future work could incorporate these more targeted methods.

Despite the model's low overall explanatory power, qualifying position remains the most influential predictor among all weekend sessions. Rather than forecasting exact outcomes, the model helps us rank which sessions contribute most meaningfully to final performance. All together, while the model indicates directional trends, the residual lack of fit may indicate that import sources of variation (e.g. weather, pit strategy and race incidents) are not fully accounted in this current model and future refinement and incorporation of additional covariates are likely to be necessary to improve calibration.

\subsubsection{Multicollinearity}

To assess the robustness of our OLR model we evaluated the presence of multicollinearity among the predictor values using Variance Inflation Factors (VIFs). High multicollinearity can distort coefficient estimates which makes it difficult to determine the true influence of each predictor on race outcomes.

We first computed the VIF scores for all predictors variables in the OLR, which were the three practice sessions (Pos\_p1, Pos\_p2, Pos\_p3) and qualification position (Pos\_q). The results can be seen in Table~\ref{tab:vif_initial} and reveal significant multicollinearity, particularly between qualification positions and the later practice sessions.

\begin{table}[h]
    \centering
    \begin{tabular}{lc}
        \toprule
        Predictor & VIF Score \\
        \midrule
        Pos\_q (Qualification Position) & 12.110 \\
        Pos\_p3 (Practice 3 Position) & 9.608 \\
        Pos\_p2 (Practice 2 Position) & 10.443 \\
        Pos\_p1 (Practice 1 Position) & 7.951 \\
        \bottomrule
    \end{tabular}
    \caption{Variance Inflation Factor (VIF) Scores (Initial Model)}
    \label{tab:vif_initial}
\end{table}

Using a baseline of a VIF score of 5 as indicating moderate collinearity, and 10+ suggesting severe, we can see from Table~\ref{tab:vif_initial} that both Pos\_q and Pos\_p2 exhibit substantial multicollinearity, with Pos\_p3 right behind.  This suggests there's a strong relationship between a driver's late weekend performance and their qualifying results which aligns with the strategy where teams are using these later sessions to fine-tune car setup and maximize qualifying performance.

Given the high collinearity among these sessions we tested two approaches to mitigate this. First we iteratively removed the highest VIF contributors, first Pos\_p2, then Pos\_p1 however this still left substantial correlation between Pos\_p3 and Pos\_q (VIF of 8.396279 for both). We then used Principal Component Analysis (PCA) to preserve predictive information by combining the three practice sessions into a single composite variable termed "Practice Performance Score" dramatically lowering VIF to near 1.000.

While PCA was mathematically effective in reducing multicollinearity, it came at a major cost to predictive power and usefulness of our model. After applying OLR we found that while the log-likelihood improved slightly (-12,802 to -12,286) and AIC dropped slightly (25,467 to 24,621) - the McFadden's $R^2$ fell to 0.00 suggesting the model lost almost all explanatory power and the Practice Performance Score became statistically insignificant ($p=0.590$).


Altogether these results confirm that qualification is the strongest individual predictor of race outcomes, but also emphasizes that Practice 3 plays a crucial role in setting up for qualification.  The strong collinearity between these sessions suggests that teams optimizing for performance in Practice 3 are more likely to achieve a strong qualifying result and thus dictate a strong race outcome.  Additional teams should be evaluating progression across the entire weekend however their combined impact cannot be captured by a single aggregate score.


\section{Discussion and Conclusion}

The results from this work provide a comprehensive overview of the relationship and evolution of results in an F1 weekend, from the three practice sessions, qualifying results and the final standings in the race itself. Prior work started looking at this relationship focusing on just starting grid positions, our work expands upon this, using a larger dataset over more seasons, to demonstrate that qualification is the strongest predictor and there is some signal with practice results. This shift emphasizes the importance of qualification results as a predictor as it's the cleanest measure of single-lap performance and speed in a week. Qualifying results, free from penalties and grid placement offer an unfiltered view of a car's performance which aids them becoming a stronger predictor of race outcome.

While qualification results are emphasized, the results from practice sessions provide key insights to teams. Practice sessions later into the weekend tend to show stronger predictive power of both the qualification and race results. This finding across multiple approaches reflects how teams do and should continue to refine their technical car setup in a weekend to achieve optimal performance with P3 becoming the final opportunity to set up for qualification. Teams looking to maximize performance and achieve the highest race results that weekend should aim to maximize extract peak performance from each session rather than use these sessions for other options (e.g. testing new drivers, testing new parts). This has to be balanced against a team's season long goals. Given the link between these sessions, if a team's goal is to maximize performance on race day, they need to try and optimize performance across all sessions consistently to be prepared for qualification.

\subsection{Future Work}


These results reinforce a concept that applies to most sports, the concept that the fastest car (or driver or athlete) will generally perform best all across all stages of a sporting event. Especially when they are performing in a process designed to measure top speed/performance. This further opens up question to understand not if the fastest car is the fastest car, but what makes a car faster. Future work should explore externals factors to determine what is making a car faster, everything from engineering and aerodynamics, team resources, reliability, weather impacts, athlete physiology and performance, and car team strategy.

Additional areas that can help build upon the foundations set up by this work. The first is the introduction of Sprint Weekends, introduced in 2021 which reduces the number of practice sessions with an additional race and qualification set up. These weekends provide fewer opportunities for teams to establish peak performance in their car and studying this limitations can help understand the impact of practice sessions in a compressed weekend and provide more actionable insights for teams.

Another area of research could be understanding the variability in races by examining reliability and how often drivers fail to finish races and the factors that contribute. Understanding frequencies and conditions in which drivers gain or lose larger places, or fail to complete the race due to mechanical issues can help place this work in a larger context and contribute to team strategy and decision making, mitigating risks and capital (especially under the newer budget cap rules). Similar, expanding upon the role of driver talent within F1, vs car mechanical/aerodynamic performance, can help understand if the limits are likely to come from the athlete side of performance, or from the engineering team.

Finally, adding additional factors to this data, where can be found (e.g. in-race telemetry extracted from F1TV data, pit stop performance, weather and altitude, sector and mini-sector results) could help determine which factors have an impact in the final race outcomes. Especially given the low $R^2$ demonstrated by our OLR, there's much room to explain the variability. These results could help better profiles which cars and drivers are able to handle different track types and know where they are likely to succeed especially compared to the baseline expectations from this presented work.

\subsection{Conclusion}

In summary this paper has reviewed the evolution of a race weekend to better understand the impacts of performance evolution and finds that qualification performance is the strongest predictor of race outcomes in an F1 race, better than what pervious work had found. The evolution of a race weekend further highlights the importance of optimizing setups through a weekend, particularly for P3 to optimize results. External factors, such as reliability, strategy, driver behaviour and weather conditions, all continue to add stochastic layers that make the final results unknown until the chequered flag is waved. Future research should explore these variables to improve the explanatory power of models and refine the theoretical understanding of how pre-race sessions interact with unpredictable race-day dynamics.

\bibliographystyle{plain}
\bibliography{references}

\begin{thebibliography}{10}

\bibitem{Bisagni2008}
C.~Bisagni and D.~Terletti.
\newblock Structural optimisation of composite elements of a formula one racing car.
\newblock {\em Journal of Composite Materials}, 42(5):471--490, 2008.
\newblock Accessed January 25, 2025.

\bibitem{FagerlandHosmer2016}
M.~W. Fagerland and D.~W. Hosmer.
\newblock Tests for goodness of fit in ordinal logistic regression models.
\newblock {\em Journal of Statistical Computation and Simulation}, 86(15):3191--3208, 2016.

\bibitem{FagerlandHosmer2017}
M.~W. Fagerland and D.~W. Hosmer.
\newblock How to test for goodness of fit in ordinal logistic regression models.
\newblock {\em Stata Journal}, 17(3):668--686, 2017.

\bibitem{FIA2024}
{Fédération Internationale de l'Automobile (FIA)}.
\newblock {Sporting Regulations for Formula One}, 2024.
\newblock Accessed January 24, 2025.

\bibitem{gofcat2023}
{Gertheiss et al.}
\newblock {\em gofcat: Goodness-of-Fit Tests for Categorical Response Regression Models}, 2023.
\newblock R package version 0.1.0.

\bibitem{Hughes1968}
D.~E. Hughes.
\newblock Measurement in motor racing performance.
\newblock {\em International Journal of Sports Science}, 8:285--292, 1968.

\bibitem{Lipsitz1996}
S.~R. Lipsitz, G.~M. Fitzmaurice, and G.~Molenberghs.
\newblock Goodness-of-fit tests for ordinal response regression models.
\newblock {\em Journal of the Royal Statistical Society: Series C (Applied Statistics)}, 45(2):175--190, 1996.

\bibitem{McCarthy2013}
L.~M. McCarthy and K.~W. Rotthoff.
\newblock Incentives on the starting grid in formula one racing.
\newblock {\em The Journal of SPORT}, 2(2):175--184, 2013.
\newblock Accessed January 25, 2025.

\bibitem{Muhlbauer2010}
T.~Mühlbauer.
\newblock Relationship between starting and finishing position in formula one car races.
\newblock {\em International Journal of Performance Analysis in Sport}, 10(2):99--105, 2010.

\bibitem{Padilla2023}
N.~Padilla.
\newblock A research study on formula one: Determining the effectiveness of rookie drivers, 2023.
\newblock Accessed January 25, 2025.

\bibitem{Schwaberger1987}
G.~Schwaberger.
\newblock Physiological demands in formula one racing: A review.
\newblock {\em British Journal of Sports Medicine}, 40(7):655--656, 1987.

\bibitem{singh_olr}
Himanshi Singh.
\newblock Multinomial and ordinal logistic regression in r, 2016.
\newblock Accessed: 2025-05-04.

\bibitem{Budden2020}
Catapult Sports.
\newblock Formula 1 race strategy analysis: The role of data in securing victory, 2023.
\newblock Accessed January 24, 2025.

\bibitem{ucla_olr}
{UCLA Statistical Consulting Group}.
\newblock R data analysis examples: Ordinal logistic regression, n.d.
\newblock Accessed: 2025-05-04.

\bibitem{VanKesteren2022}
E.-J. van Kesteren and T.~Bergkamp.
\newblock Bayesian analysis of formula one race results: Disentangling driver skill and constructor advantage.
\newblock {\em Journal of Quantitative Analysis in Sports}, 18(2):85--104, 2022.

\end{thebibliography}

\end{document}